\documentclass[letter]{aa}
\usepackage[varg]{txfonts}
\usepackage{graphicx}
\RequirePackage[colorlinks=true,linkcolor=blue,citecolor=blue,urlcolor=blue]{hyperref}

\newcommand{\dcdot}{\mathbin{%
    \nonscript\mspace{-\muexpr\medmuskip*2/3}%
    \cdot
    \nonscript\mspace{-\muexpr\medmuskip*2/3}%
  }%
}

\begin{document}

\title{Zooming in on the Circumgalactic Medium with GIBLE: \\the Topology and Draping of Magnetic Fields around Cold Clouds}

\author{Rahul Ramesh\inst{1}\thanks{E-mail: rahul.ramesh@stud.uni-heidelberg.de}
\and Dylan Nelson\inst{1}
\and Drummond Fielding\inst{2,3}
\and Marcus Br\"{u}ggen\inst{4}
}
\institute{
Universität Heidelberg, Zentrum für Astronomie, ITA, Albert-Ueberle-Str. 2, 69120 Heidelberg, Germany \label{1}
\and Center for Computational Astrophysics, Flatiron Institute, 162 Fifth Avenue, New York, NY 10010, USA \label{2}
\and Department of Astronomy, Cornell University, Ithaca, NY 14853, USA \label{3}
\and Hamburg University, Hamburger Sternwarte, Gojenbergsweg 112, 21029 Hamburg, Germany \label{4}
}

\date{}

\abstract{We use a cosmological zoom-in simulation of a Milky Way-like galaxy to study and quantify the topology of magnetic field lines around cold gas clouds in the circumgalactic medium (CGM). This simulation is a new addition to Project GIBLE, a suite of cosmological magnetohydrodynamical simulations of galaxy formation with preferential super-Lagrangian refinement in the CGM, reaching an unprecedented (CGM) gas mass resolution of $\sim$\,$225$\,M$_\odot$. To maximize statistics and resolution, we focus on a sample of $\sim$\,$200$ clouds with masses of $\sim$\,$10^6$\,M$_\odot$. The topology of magnetic field lines around clouds is diverse, from threading to draping, and there is large variation in the magnetic curvature ($\kappa$) within cloud-background interfaces. We typically find little variation of $\kappa$ between upstream and downstream cloud faces, implying that strongly draped configurations are rare. In addition, $\kappa$ correlates strongly with multiple properties of the interface and the ambient background, including cloud overdensity and relative velocity, suggesting that cloud properties impact the topology of interface magnetic fields.}


\titlerunning{Topology and Draping of Magnetic Fields around Cold Clouds with GIBLE}
\authorrunning{R. Ramesh et al.}

\maketitle

\section{Introduction}

Both observations and simulations suggest that galaxies are surrounded by a multi-phase, multi-scale reservoir of gas. Termed the circumgalactic medium (CGM), this gaseous halo is believed to play a vital role in the growth and evolution of galaxies (see \citealt{donahue2022} for a recent review of the CGM). While the volume of the CGM is dominated by a warm-hot component, it can also host small, cold gas structures. The high-velocity clouds (HVCs) of the Milky Way are prototypical examples \citep[e.g.][]{muller1963,wakker1997}.

Despite having been first observed several decades ago, there remain a number of open questions regarding HVCs, and cold CGM clouds in general. Their expected lifetimes, and the nature of their growth and evolution, are uncertain. A number of idealised `cloud-crushing' simulations have explored these puzzles. While early studies suggested that cloud lifespans should be short \citep[e.g][]{klein1994,mellema2002}, certain physical mechanisms could enhance their survival. For instance, the Kelvin-Helmholtz instability may produce a warm interface layer between the cold cloud and the hot background, facilitating rapid cooling and cloud growth \citep[e.g.][]{scannapieco2015,gronke2018,fielding2020}. 

In addition, non-thermal components including magnetic fields may be important. They can suppress fluid instabilities \citep[e.g.][]{berlok2019,sparre2020,das2024}, provide non-thermal pressure support \citep[e.g.][]{girichidis2021,hidalgo2023,fielding2023} or enhance the Rayleigh-Taylor instability, accelerating condensation \citep{gronnow2022}.
The direction and topology of magnetic field lines may further be important by influencing the amplification of magnetic energy density \citep{shin2008}, kinematics \citep{kwak2009} and shape \citep{banda2016,bruggen2023} of clouds. 

While these theoretical studies have advanced our understanding of cloud growth and survival, they have a fundamental limitation -- they are all idealised, non-cosmological simulations. As a result, they must assume the existence of a pre-existing cloud, and background, with particular properties. In the case of magnetic fields, the strength and orientation must be chosen ad hoc, i.e. freely explored. Cosmological simulations overcome this limitation by self-consistently evolving halo gas and magnetic fields over cosmic epochs, with the trade-off of coarser resolution. Recent cosmological simulations including TNG50 have been shown to realise small-scale cold gas structures \citep{nelson2020,ramesh2023b}, even at the limited resolution available in large uniform volumes.

We here take a step forward by using Project GIBLE \citep{ramesh2023d}, a suite of cosmological zoom-in galaxy formation simulations with targeted, additional super-Lagrangian refinement of gas in the CGM. In particular, we present a new simulation of a Milky Way-like galaxy run to $z=0$ with even higher resolution than our first GIBLE results. These simulations make it possible to better resolve and study small-scale phenomena in the full $\Lambda$CDM cosmological context, thereby bridging the gap between highly resolved idealised simulations, and more realistic cosmological runs at lower resolution. 
Building on our earlier work on the magnetothermal properties of the clumpy CGM in a cosmological context, we now, for the first time, quantify the topology and draping of magnetic field lines around cold, dense gas clouds in a self-consistent environment without the sensitivity to initial magnetic field geometries that limit the robustness of previous studies on this topic due to their idealized nature.

The paper is structured as follows: in Section~\ref{sec:methods} we describe Project GIBLE and our methodology. Results are presented in Section~\ref{sec:results}, discussed in Section~\ref{sec:disc} and summarised in Section~\ref{sec:summary}. 

\section{Methods}\label{sec:methods}

\subsection{Simulation Overview}

In this paper we use Project GIBLE \citep{ramesh2023d}, a suite of cosmological magneto-hydrodynamical zoom-in simulations of Milky Way-like galaxies ($M_\star \sim 10^{10.9}$\,M$_\odot$, $M_{\rm{200c}} \sim 10^{12.2}$\,M$_\odot$). In particular, we present a new `RF4096' simulation, of a single halo, with preferential mass refinement that achieves a gas mass resolution of $\sim$\,$225$\,M$_\odot$ in the CGM, defined as the region bounded between 0.15\,R$_{\rm{200c}}$ and R$_{\rm{200c}}$ (virial radius). This is the latest addition to Project GIBLE, currently comprised of eight Milky Way-like galaxies each simulated at CGM gas mass resolutions of $\sim$\,$10^3$, $10^4$ and $10^5$ M$_\odot$, labelled the `RF512', `RF64' and `RF8' suites, respectively. In all cases, the galaxy is maintained at a resolution of $\sim$\,$8.5 \times 10^5$\,M$_\odot$.

Project GIBLE uses the IllustrisTNG model \citep{weinberger2017,pillepich2018}, within the \textsc{arepo} code \citep{springel2010}, to account for the physical processes that regulate galaxy formation and evolution. This includes radiative thermochemisty and metal cooling with the metagalactic radiation field, star formation, stellar evolution and enrichment, supermassive black hole (SMBH) formation, and feedback from stars (supernovae) and SMBHs (i.e. AGN; thermal, kinetic, and radiative modes).

The TNG model also includes ideal magneto-hydrodynamics \citep{pakmor2014} -- a uniform primordial field of $10^{-14}$ comoving Gauss is seeded at the start of the simulation, which is subsequently (self-consistently) amplified as a combined result of structure formation, small-scale dynamos and feedback processes \citep{pakmor2020}. While the initial field is divergence free by construction, the \cite{powell1999} cleaning scheme is used to maintain $\nabla \dcdot \vec{B} = 0$ over time. We note that the relative divergence error is typically small ($\lesssim O(10^{-2})$), indicating that divergence errors are minimal \citep[see also][]{pakmor2013}. The order of magnitude of field strengths predicted to be found in the gaseous halos around galaxies by the TNG model \citep{marinacci2018, nelson2018, ramesh2023c, ramesh2023a} is broadly consistent with recent indications for the existence of large-scale $\sim$\,$\rm{\mu G}$ magnetic fields in the CGM observed using Faraday rotation \citep{heesen23, boeckmann23}.

\subsection{Cloud and Interface Identification Algorithm}\label{ssec:cloud_id}

Following \cite{nelson2020,ramesh2023b}, we define and identify clouds as spatially contiguous sets of cold ($T \leq 10^{4.5}$\,K) Voronoi cells, i.e. collections of cold gas cells that are Voronoi natural neighbors. Further, we consider only those gas cells that are not gravitationally bound to any satellite galaxies, as identified by the substructure identification algorithm \textsc{subfind} \citep{springel2001}. To maximize statistics and resolution, throughout this work, we restrict the selection of clouds to those with masses in the range $[10^{5.8}, 10^{6.2}]$\,M$_\odot$, resulting in a sample size of $233$. These clouds, on average, are resolved by $\sim$\,$3700$ gas cells, with more in the surrounding interfaces.

The interface layer of each cloud is defined as all gas cells that are not cold ($T > 10^{4.5}$\,K) but share a face with a member cell of the cloud, determined using the Voronoi tessellation connectivity\footnote{While interface gas cells are typically contiguous amongst themselves, there are rare cases where `gaps' may be present in the interface.}. This interface layer is thus the cocoon of cells immediately surrounding the cloud. On average, the interface layers around our clouds are sampled by $\sim$\,$2550$ resolution elements.

We quantify the topology of magnetic field lines in the interface layer using measurements of magnetic curvature ($\kappa$), defined as \citep{shen2003,pfrommer2022}:
\begin{equation}
    \kappa = |\Vec{\kappa}| = |(\Vec{b} \dcdot \nabla) \Vec{b}|
\label{eq:kappa}    
\end{equation}
where $\Vec{b}$ = $\Vec{B} / B$ is the unit vector in the direction of the magnetic field $\Vec{B}$. Vector $\Vec{\kappa}$ points in the direction of the local centre of curvature of $\Vec{B}$, while $\kappa$ corresponds to the inverse of the radius of curvature \citep{boozer2005}. A value of $\kappa=0$ thus describes a straight field, while larger values denote field lines with increasing deviations from uniformity.

Following a calculation of $\kappa_{\rm{cell}}$ for each interface gas cell using Equation~\ref{eq:kappa}, we compute the magnetic curvature around a cloud as the mean of all its interface gas cells. We denote this mean magnetic curvature as $\kappa$ throughout the rest of the text.

\begin{figure}
    \centering
    \includegraphics[width=9cm]{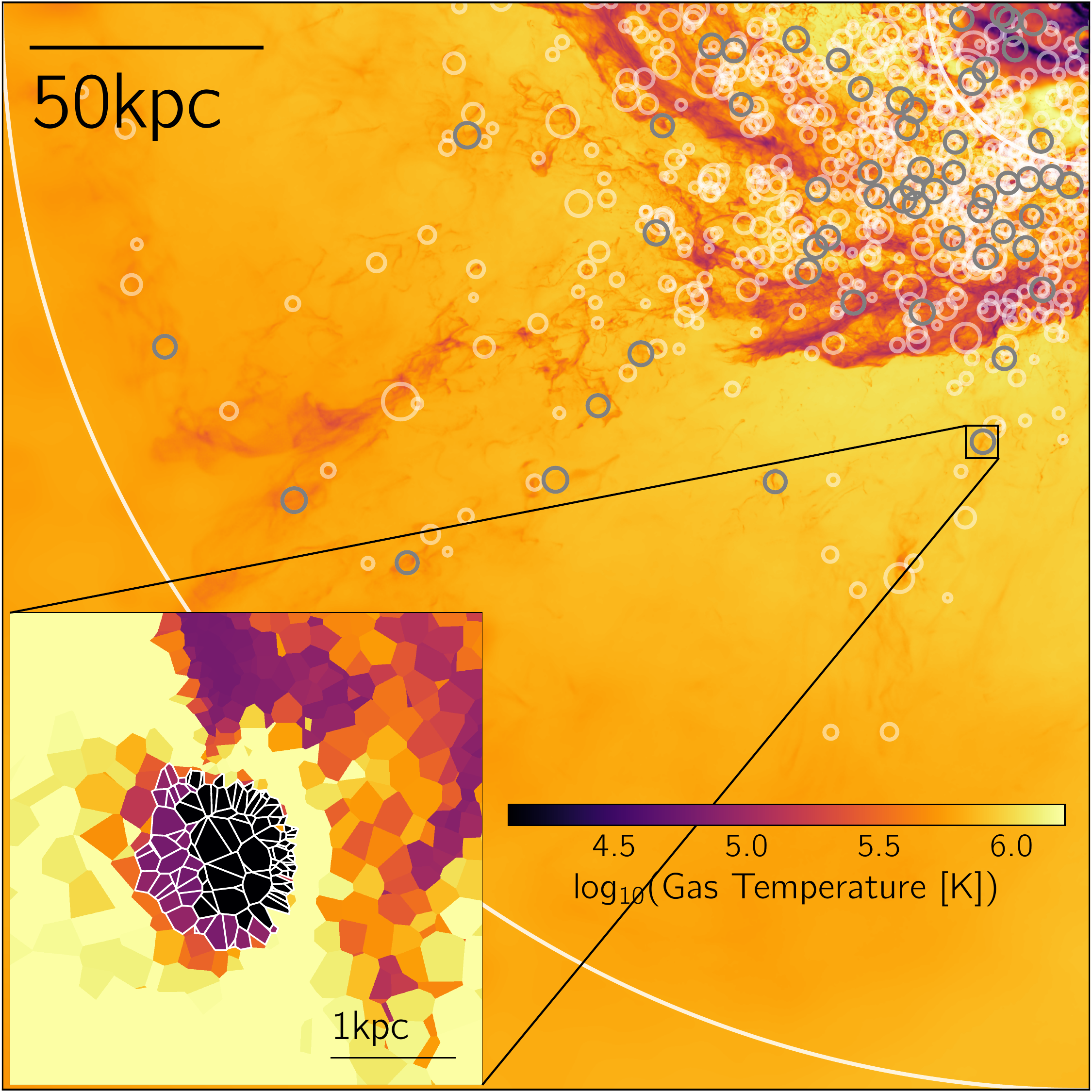}
    \caption{A visualisation of the distribution of clouds in a quadrant of our highest resolved GIBLE halo, a Milky Way-like galaxy at $z=0$. The centre of the galaxy is in the top-right corner (main image). The image extends $\rm{R_{200c}}$ from edge-to-edge, and $\pm \rm{R_{200c}}$ along the projection direction. Colors show mean mass-weighted gas temperature in projection. Circles show the positions of the many hundred cold, dense CGM gas clouds with masses greater than $10^5$\,M$_\odot$. Our fiducial sample with $M_{\rm{cl}} \sim$\,$10^6$\,M$_\odot$ is marked in gray. The inset, a highly zoomed-in region of the halo, shows a slice of the Voronoi mesh centred around a random cloud from our sample, with all member cells outlined by white lines. Despite their small sizes, we resolve such clouds (and their interface layers) with $\sim$\,$3700$ ($2550$) gas cells, enabling us to study small-scale phenomena self-consistently evolved in a cosmological context.}
    \label{fig:introVis}
\end{figure}

\begin{figure*}
    \centering
    \includegraphics[width=8.5cm]{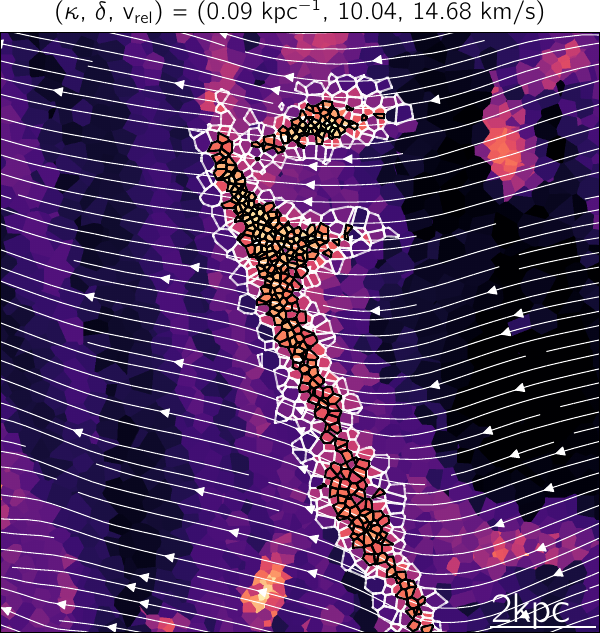}
    \includegraphics[width=8.5cm]{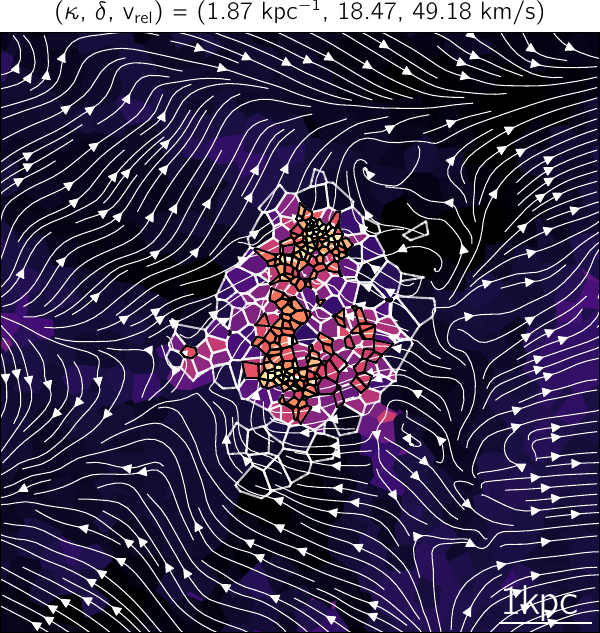}
    \includegraphics[width=6cm]{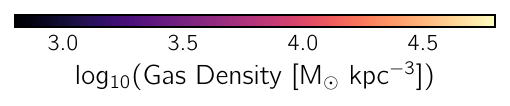}
    \caption{A visualisation of the diverse topologies of magnetic field lines around cold CGM clouds. The two panels show slices of the Voronoi mesh centred around two clouds, both oriented such that their mean velocity vector i.e. direction of motion are to the right, the positive x-axis direction. Cells that belong to the cloud (interface layer) are outlined using black (white) lines. Streamlines show the direction of magnetic field lines. The three numbers on the top of each image correspond to the mean magnetic curvature of the cloud interface ($\kappa$), as well as the overdensity ($\delta$) and relative velocity ($\rm{v_{rel}}$) between the cloud and the interface layer. While the left cloud is threaded by magnetic fields, in a region of the background CGM which has particularly uniform field orientation, the magnetic fields in the right panel begin to respond to the motion of the cloud. This diversity is captured by the different values of $\kappa$.}
    \label{fig:cloudVis}
\end{figure*}

\section{Results}\label{sec:results}

We begin with a visualisation of the distribution of clouds in the CGM of our simulated halo in Fig.~\ref{fig:introVis}. The image, which shows one quadrant of the halo at $z=0$\footnote{We exclusively consider the $z=0$ simulation snapshot to best connect with the observational Milky Way HVC community.}, extends $\rm{R_{200c}}$ ($\sim$\,236\,kpc) from edge-to-edge, and $\pm \rm{R_{200c}}$ along the projection direction, with the centre of the galaxy located at the top-right corner. Colors show the average (mass-weighted) temperature of gas along the line of sight. The small white circles mark the positions of all clouds with masses greater than $10^5$\,M$_\odot$, with radii scaling with the mass of clouds. Of these many hundreds of clouds, the fiducial sample that we consider in this work, $10^{5.8} < M_{\rm cl} / \rm{M_\odot} < 10^{6.2}$, is shown in gray. These cool clouds are embedded in the $\sim 100$ times hotter, volume filling background CGM.

The inset shows a slice of the Voronoi mesh centred around a single cloud. We outline cells belonging to the cloud with white lines. The ratio of the physical scales of the two images is $\sim$\,$60$, i.e. the inset shows a highly zoomed-in region of the main image, but is still well resolved. Simulations of the kind shown here thus enable the study of small-scale cloud phenomena, including formation, evolution, mixing, and so on, with clouds self-consistently evolved in the full cosmological context.

In Fig.~\ref{fig:cloudVis}, we show two examples of the topology of magnetic field lines around cool clouds. The two panels show slices of the Voronoi mesh centred around two distinct clouds. Both are oriented such that their velocities, computed as the mass-weighted mean velocity of all cloud member cells, point along the positive $x$-axis. Both clouds are infalling towards the centre of the galaxy, and are at similar galactocentric distances ($\sim$\,$50$\,kpc). Cells that belong to the clouds (interface layers) are outlined with black (white) lines. Streamlines show the direction of magnetic field lines in this plane, while background color corresponds to the density of gas. We include three numbers atop each figure: the magnetic curvature averaged over all interface gas cells ($\kappa$), the ratio of the mean density of the cloud to that of the ambient background\footnote{We define the ambient background to be comprised of three layers of non-cold gas cells around clouds.}, i.e. the density contrast ($\delta$), and the modulus of the difference between the cloud velocity and that of the ambient background, i.e. the velocity contrast ($\rm{v_{rel}}$).

\begin{figure}
    \centering
    \includegraphics[width=8.5cm]{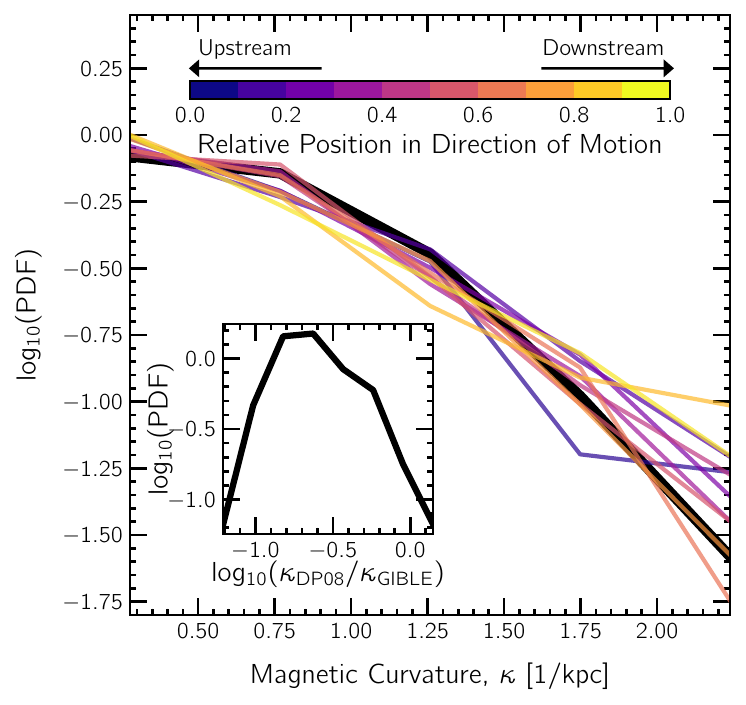}
    \caption{Distribution of interface magnetic curvature values for our sample of $\rm{M_{cl}} \sim$\,$10^6$\,M$_\odot$ clouds (main panel). The black curve is based on all interface gas, while the other curves show values derived using gas with different relative positions with respect to the direction of motion of the cloud. Purple curves show $\kappa$ for the head/upstream regions, while yellow curves show $\kappa$ in the tail/downstream regions. The inset compares the upstream interface magnetic curvature from our simulations ($\kappa_{\rm{GIBLE}}$) to a simple theoretical model ($\kappa_{\rm{DP08}}$).}
    \label{fig:mc_pdf}
\end{figure}

\begin{figure*}
    \centering
    \includegraphics[width=18.25cm]{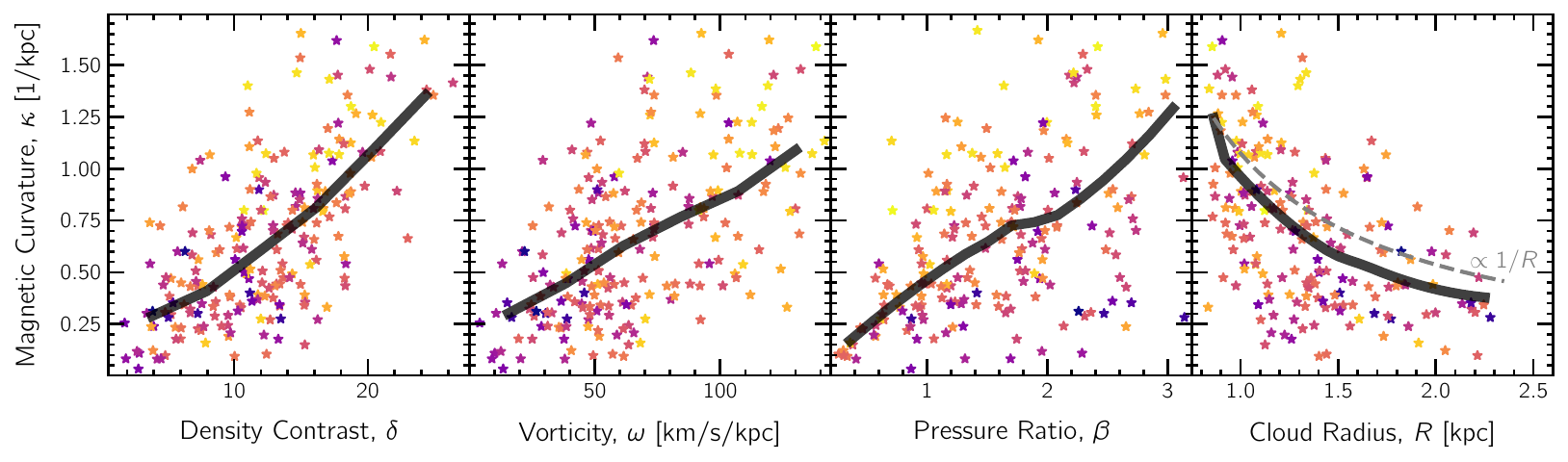}
    \includegraphics[width=6cm]{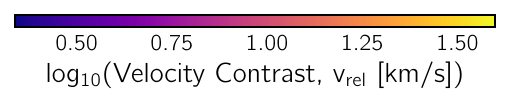}
    \caption{From left to right, magnetic curvature ($\kappa$) as a function of density contrast between the cloud and its ambient background, vorticity in the interface layer, thermal-to-magnetic pressure ratio of the interface, and cloud radius. The solid curves show the median, while scatter points correspond to individual clouds, each colored by the velocity contrast between the cloud and its ambient background. A strong trend of $\kappa$ is seen in the median with respect to each of the properties considered, while the variation of $\kappa$ at fixed abscissa is typically well captured by the diversity of velocity contrasts of clouds.}
    \label{fig:mc_v_dc}
\end{figure*}

The magnetic field lines around these clouds show contrasting structures. While the field is largely coherent in the left panel, quantified by a relatively low $\kappa$ of $\sim$\,0.1\,kpc$^{-1}$, the topology is more complex on the right with tangled and less laminar field lines ($\kappa$\,$\sim$\,1.87\,kpc$^{-1}$). The numbers atop each panel show that $\kappa$ correlates with, amongst other properties, contrasts in both density and velocity, which we consider further in Figure~\ref{fig:mc_v_dc}.

In the main panel of Fig.~\ref{fig:mc_pdf}, we explore the variation in values of $\kappa$ across our sample. The black curve corresponds to the computation of $\kappa$ as the mean over all interface gas cells, i.e. our fiducial definition. The other curves instead compute averages of a sub-selection of interface gas cells from $10$ bins, constructed based on their relative position in the direction of motion of the cloud, as shown by the colorbar. A relative position of $0.1$ would thus correspond to the first $10\%$ of interface gas cells in the direction of motion, $0.2$ to the next $10\%$, and so on. 

The black distribution peaks at a value of $\kappa$ of $\lesssim$\,$0.1$\,kpc$^{-1}$, is relatively flat out to $\kappa$ $\sim 0.7$\,kpc$^{-1}$, and drops steadily towards larger values. A significant fraction of clouds are thus predicted to have largely coherent fields around them in their interface regions, much like in the left panel of Fig.~\ref{fig:cloudVis}. The colored curves show similar behavior, with little dependence on relative position. That is, the degree of curvature does not strongly change between the upstream and downstream interface regions. Previous studies with idealised simulations have shown that field line draping around clouds moving through an (initially) uniform magnetic field perpendicular to the motion of the cloud is more (less) efficient in the head (tail) \citep[e.g.][]{jung2023}, corresponding to larger (smaller) values of $\kappa$ upstream (downstream) of the cloud. The insignificant difference in the distributions of $\kappa$ between these regions in Fig.~\ref{fig:mc_pdf} suggests that such strong draping configurations are not common around our simulated clouds. We speculate that this is largely due to background field lines upstream of the cloud not being oriented in perpendicular directions and as uniform as is typically assumed in idealised setups \citep[see also][]{sparre2020}. We note here that distributions explored in Fig.~\ref{fig:mc_pdf} are largely converged up to our `RF512' simulations, i.e. 8 times lower mass resolution (not shown explicitly).

As the value of $\kappa$ corresponds to the inverse of the radius of curvature of the local magnetic field, it should scale inversely with the radius of the cloud for strongly draped configurations. However, we have checked that the above results are qualitatively similar when values of $\kappa$ are normalised\footnote{Following \cite{nelson2020}, we define the effective radius by the volume equivalent sphere, $R = [3 V_{\rm{cloud}} / 4 \pi]^{1/3}$.} by $1/R$ (see also the large diversity of $\kappa$ at fixed $R$ in Fig.~\ref{fig:mc_v_dc}). The $\kappa$ distribution therefore reflects physically different field geometries in the cloud interfaces.

In the inset, we make a comparison to a simple model describing the expected field around a sphere of radius $R$ moving through a homogeneous ambient medium with an (initial) uniform magnetic field oriented perpendicular to the cloud motion \citep{dursi2008}. Given the assumptions of this model, the solution is not valid in the wake behind the sphere. We thus restrict our comparison to the region around the head of the cloud, which we define using the angular bounds $\theta=[-\pi/3,\pi/3]$ and $\phi=[-\pi/3,\pi/3]$. Although this choice is arbitrary, we note that adopting other angular ranges for $\theta$ and $\phi$ has no significant impact on the analysis that follows. We compute the theoretical estimate of the model, $\kappa_{\rm{DP08}}$, for each cloud separately by setting $R$ to the effective radius of the cloud.

The curve shows the distribution of the ratio $\kappa_{\rm{DP08}} / \kappa_{\rm{GIBLE}}$. The PDF peaks around $-0.7$, with a large spread, and the model typically under-predicts the curvature seen in the simulation. Although not shown here, we find a weak anti-correlation in this ratio with $\delta$ and v$_{\rm{rel}}$\footnote{Note that the value of $\kappa$ predicted by the \cite{dursi2008} model only depends on $R$, and does not take $\delta$ and v$_{\rm{rel}}$ as input parameters.}, suggesting that the model works better for clouds with relatively small density and velocity contrasts. We speculate that this could be linked to the more efficient build-up of magnetic fields around objects of greater over-densities and velocity contrasts \citep{lyutikov2006}, thereby increasing the impact of magnetic back-reaction on the flow, which is not considered in the \cite{dursi2008} model.

Finally, we return to the correlation between $\kappa$ and properties of the interface and the ambient background. From left to right, Fig.~\ref{fig:mc_v_dc} shows $\kappa$ as a function of the density contrast ($\delta$), vorticity in the interface layer ($\omega = \nabla \times v$), the thermal-to-magnetic pressure ratio of the interface ($\beta$), and the radius of the cloud ($R$). The solid curve shows the median, while colored points correspond to individual clouds, each colored by v$_{\rm{rel}}$. On average, $\kappa$ increases almost linearly with $\delta$. The scatter clearly correlates with relative velocity: larger values of $\kappa$ have higher v$_{\rm{rel}}$ at fixed $\delta$. A linear correlation is also present between $\kappa$ and $\omega$, indicating a possible connection between a disordered velocity field and a disordered magnetic field. Consistent with theoretical predictions for the case of an (initial) magnetic field
that is coherent on scales larger than the cloud size \citep{mccourt2015}, we find that $\kappa$ correlates with $\beta$. A least-squares fit yields $\kappa \propto \beta^{0.9}$, roughly in the ballpark of the predicted $\kappa \propto \beta$ trend by \cite{schekochihin2004} for the case of small-scale turbulent dynamos. While $\kappa$ decreases with increasing $R$, the drop is steeper than the $1/R$ trend expected for a strongly draped configuration (see discussion above), suggesting again that such configurations are not common in our sample. Similar to before, we find that results shown here are largely converged up to our `RF512' level runs.

\section{Discussion}\label{sec:disc}

The sizes, density contrasts and kinematics of clouds may thus have an important impact on the structure of ambient field lines. The resulting magnetic field topology may in turn affect cloud growth and evolution. For example, the draped magnetic field layer increases the drag force by a factor of $\sim$\,$[1 + (v_{\rm{A}} / v_{\rm{rel}})^2] \equiv [1 + (R \kappa)^{-1}]$, where $v_{\rm{A}}$ is the Alfven speed in the background. The enhanced drag force decreases the `stopping distance' by $[1 + (R \kappa)^{-1}]^{-1}$, i.e. the distance travelled by the cloud prior to achieving velocity equilibrium with its surroundings, thereby improving chances of their survivability \citep{mccourt2015}. The diversity of cloud and interface properties portrayed by Figure~\ref{fig:mc_v_dc} suggests that the impact of field line topology on a population of clouds is expected to be varied. Specifically, at fixed $R$,  clouds with lower $\delta$/$\beta$/$v_{\rm{rel}}$ typically have lower values of $R \kappa$, and would thus experience a larger boost in their drag force as compared to high $\delta$/$\beta$/$v_{\rm{rel}}$ counterparts\footnote{Note that this only describes the enhancement factor of the drag force as a result of draping. The total drag force experienced by the cloud ($\sim$\,$\rho_{\rm{interface}}^2 v_{\rm{rel}}^2 R^2 [1 + (R \kappa)^{-1}]$) depends on other properties of the cloud and the interface as well.}. We reiterate that cosmological simulations like GIBLE enable us to assess such predictions for an actual, diverse cloud population, since clouds, their interfaces and magnetic fields evolve self-consistently.

The magnetic field topology and draped layers may furthermore play a role in suppressing the impact of the Kelvin-Helmholtz instability along the surface of the cloud \citep[e.g.][]{pfrommer2010}. For instance, at fixed $B_{\rm{wind}}$, \citealt{sparre2020} showed that clouds with draped topologies in their interfaces are expected to survive longer. In addition, \citealt{jung2023} find that regions where field lines are inefficiently draped (i.e. low values of $\kappa$) fragment rapidly into smaller clumps, while regions of high $\kappa$ are instead extended into long filamentary structures as a result of enhanced magnetic tension and effectively survive longer. However, it is important to note that the net impact of this suppression of mixing on the evolution of clouds may depend on the efficiency of radiative cooling in the interface layer \citep[e.g.][]{gronke2018}. In agreement with idealized work, clouds in TNG50 have temperature gradients into their interfaces, i.e. they are surrounded by a mixed-phase layer of warm gas that rapidly cools onto the cloud \citep{nelson2020}, consistent with theoretical local cooling flow models \citep{dutta22}. Moreover, the metal content of clouds and their interfaces can vary significantly \citep{nelson2020,ramesh2023b}, possibly affecting the rate at which gas in the interface condenses. Future work will quantify the resulting impact on cloud growth and survival in our simulations with Lagrangian tracers (\textcolor{blue}{Ramesh et al. in prep}). 

While we find that clouds are roughly in pressure balance with their interface layers \citep{ramesh2023b}, thereby preventing them from being crushed and dissolved, the TNG model does not include thermal conduction. The inclusion of this component may contribute to cloud evaporation \citep[e.g.][]{marcolini2005,vieser2007}, although certain configurations of magnetic fields may partially suppress this effect \citep{ettori2000,bruggen2023}. Future simulations that include conduction can explore its role in cosmological cloud evolution. This will require that we adequately resolve the Field length \citep{field1965} to avoid spurious numerical effects \citep{koyama2004}. For example, for 10\% Spitzer conduction \citep[see e.g.][]{bruggen2023}, the Field length would be $\sim$\,$120$\,pc for interface gas cells, requiring a spatial resolution of $\lesssim 40$\,pc in this region\footnote{This `Field condition', that spatial resolution is better than the Field length by at least a factor of 3, was derived using one-dimensional simulations with isotropic conduction \citep{koyama2004}.}, i.e. $2-4 \times$ better spatial resolution than our current RF4096 run \citep[see Figure 2 of][]{ramesh2023d}.

\section{Summary}\label{sec:summary}

In this paper, we have used a cosmological zoom-in galaxy formation simulation with additional CGM refinement to study and explore the complex and diverse topology of magnetic field lines around cold, dense clouds in the CGM. At an average baryonic mass resolution of $\sim$\,$225$\,M$_\odot$, the interface layers around our sample of $10^{5.8} < M_{\rm cl} / \rm{M}_\odot < 10^{6.2}$ clouds are resolved by over 2000 resolution elements, allowing the study of interface phenomena in a cosmological context.

We quantify the structure of magnetic field lines around clouds, i.e. in interface layers, by the magnetic curvature $\kappa$. We find that values of $\kappa$ vary significantly, reflecting the diversity in field line topologies around clouds. There is no significant difference in the distribution of $\kappa$ between the regions upstream and downstream of the cloud, suggesting that strong draping configurations are rare in our sample. However, curvature correlates strongly with cloud-background contrasts in density and velocity: larger contrasts correspond to larger $\kappa$, on average. In addition, $\kappa$ also correlates with other interface properties, including vorticity and the thermal-to-magnetic pressure ratio.

This study provides a first perspective from the point of view of cosmological simulation regarding the topology of magnetic field lines around cold clouds. However, there are several clear avenues to extend this work. In particular, we can assess the impact of cloud motion on the immediate interface layer, as well as on the broader local gaseous environment of clouds. With Lagrangian tracers we can also quantify the impact of magnetic fields on the lifetime, survival, and evolution of clouds.

\begin{acknowledgements}
RR and DN acknowledge funding from the Deutsche Forschungsgemeinschaft (DFG) through an Emmy Noether Research Group (grant number NE 2441/1-1). RR is a Fellow of the International Max Planck Research School for Astronomy and Cosmic Physics at the University of Heidelberg (IMPRS-HD).
MB acknowledges support from the Deutsche Forschungsgemeinschaft under Germany's Excellence Strategy - EXC 2121 "Quantum Universe" - 390833306 and from the BMBF ErUM-Pro grant 05A2023.
This work has made use of the VERA supercomputer of the Max Planck Institute for Astronomy (MPIA), and the COBRA supercomputer, both operated by the Max Planck Computational Data Facility (MPCDF), and of NASA's Astrophysics Data System Bibliographic Services. 
\end{acknowledgements}

\bibliographystyle{aa}
\bibliography{references}

\end{document}